\documentclass[preprint,5p,times]{elsarticle}
\usepackage{lineno}
\usepackage[T1]{fontenc}
\usepackage[utf8]{inputenc}
\usepackage{graphicx}
\usepackage{textcomp}
\journal{NIM A}

\begin{document}
\begin{frontmatter}
\title{Tests of Modulated Intensity Small Angle Scattering
  in time of flight mode}
\author[frm,e21]{G. Brandl}
\author[arg]{J. Lal}
\author[arg,ornl]{J. Carpenter}
\author[ornl]{L. Crow}
\author[ornl]{L. Robertson}
\author[frm,e21]{R. Georgii}
\ead{Robert.Georgii@frm2.tum.de}
\author[e21]{P. Böni}
\author[delft]{M. Bleuel}
\address[frm]{Forschungsneutronenquelle Heinz Maier-Leibnitz,
  Technische Universität München, Lichtenbergstr.\ 1, 85747 Garching, Germany}
\address[e21]{Physik Department E21,  Technische Universität München,
  James-Franck-Str., 85747 Garching, Germany}
\address[arg]{Argonne National Laboratory, Materials Science Division,
  Argonne, IL 60439, USA}
\address[ornl]{Oak Ridge National Laboratory, Oak Ridge, TN 37831-6477, USA}
\address[delft]{Technical University of Delft, Mekelweg 15, 2629JB Delft,
  Netherlands}
\date{\today}

\begin{abstract}
  We report results of tests of the MISANS technique at the CG-1D beamline at
  High Flux Isotope Reactor (HFIR), Oak Ridge National Laboratory (ORNL).  A
  chopper at 40\,Hz simulated a pulsed neutron source at the beamline.  A compact
  turn-key MISANS module operating with the pulsed beam was installed and a well
  characterised MnSi sample was tested. The feasibility of application of high
  magnetic fields at the sample position was also explored. These tests
  demonstrate the great potential of this technique, in particular for examining
  magnetic and depolarizing samples, under extreme sample environments at pulsed
  sources, such as the Spallation Neutron Source (SNS) or the planned European
  Spallation Source (ESS).
\end{abstract}
\end{frontmatter}

\section{Introduction}
MISANS, MIEZE (Modulation of Intensity with Zero Effort) in the Small Angle
Neutron Scattering (SANS) geometry is a rather new technique to probe
quasi-elastic scattering with extremely high energy resolution. The method is
well understood \cite{Gaehler:89,Hank:97,Keller:93} and efforts are under way
\cite{Georgii:11,Brandl:11} to establish the technique as a standard tool for
measurements of slow dynamics.

The general trend of new neutron sources to be accelerator driven and thus to
provide pulsed neutron beams raises the question how MISANS will perform in a
pulsed mode. In earlier experiments the feasibility of MISANS on pulsed sources
was demonstrated \cite{Bleuel:06,Hayashida:07}, however these tests were only
using the direct beam and relative low MIEZE frequencies.

The goal of this experiment was to show that a MIEZE can be set up easily at a
new beamline and works well in the time-of-flight mode with samples. Therefore a
compact turn-key MISANS setup from the FRM II in Munich \cite{Georgii:11} was
installed at the HFIR in Oak Ridge, Tennessee, USA, at the beamline CG-1D \cite{CG1}.

\section{Setup of the Experiment}

\begin{figure}[h]
  \begin{center}
    \includegraphics[width=\linewidth]{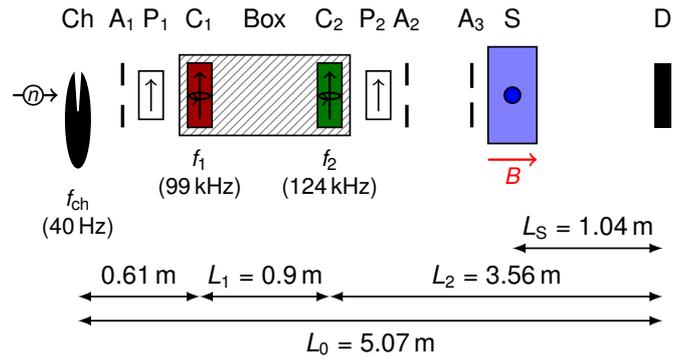}
  \end{center}
  \caption{\label{fig:setup}
    Sketch of the setup used for the MISANS experiment during the second beam
    session, viewed from the side.
    The chopper (Ch) provided a pulsed beam with a frequency of 40\,Hz and a
    pulse length of 0.14\,ms, corresponding to a chopper opening of
    $2^\circ$.  The two polarizers (P$_1$ and P$_2$) were placed before and
    after the zero field region (hatched) of the MIEZE box, which contained
    the two MIEZE coils (C$_1$ and C$_2$).  Three apertures (A$_1$ to A$_3$)
    were installed so that the beam is restricted to the sample (S), located
    inside a cryomagnet.  The time-resolved detector (D) was placed at the
    correct distance to fulfill the MIEZE condition (eq. \ref{eq:miezecond}).    
}
\end{figure}

The used setup is shown in Figure \ref{fig:setup}: Two sets of MIEZE coils
inside \textmu-metal shielding \cite{Arend:04,Georgii:11} are placed between two
polarizers.  The polarizers are polarising solid state benders \cite{Krist:02},
which polarise the transmitted beam, but keep the collimation intact
\cite{Bleuel:07}.  We used two approximately $10\times10$\,mm$^2$ apertures
(A$_1$ and A$_3$ in Figure \ref{fig:setup}) in about 3\,m distance from each
other for a rough collimation and suppression of the reflected spin states from
the polarizers.  The frequencies in the RF coils ranged between 46 and 149\,kHz
and matched the static fields:
\begin{equation}
  \omega = 2\pi \cdot f = \gamma B \quad\mathrm{with}\quad 
  \gamma = 2\pi \cdot 2.913 \,\mathrm{kHz/G}.
\end{equation}

In order to simplify the setup, the coupling coils were replaced by small guide
fields and therefore all static fields were either parallel or anti-parallel to
the strong magnetisation fields of the polarizers.  This simplification of the
setup removed the need for an adiabatic rotation of the neutron polarisation in
the static fields by 90$^\circ$ as usual in MIEZE experiments and thus
allowed a quicker adjustment of the setup in a new environment.  The trade-off
is that the MIEZE coils can only be operated in a $\pi/2$-mode, since the
initial polarization is parallel to the static field of the $B_0$ coils.  In
this mode, each MIEZE coil must perform a $\pi/2$ flip to achieve the coherent
splitting of two spin states that leads to the focused coherent overlap at the
MIEZE point \cite{Ebisawa:04}.

In contrast, in the usual MIEZE setup, which is preferred for all new
developments, each MIEZE coil induces a $\pi$ flip, and in combination with a
``bootstrap'' arrangement a factor of four in time resolution is gained in
comparison to our $\pi$/2 mode setup.

For this time-of-flight MIEZE instrument, the RF amplitude needs to be modulated
to match the condition
\begin{equation}
  \label{eq:rfmod}
  \gamma B \frac{l}{v} = \frac{\pi}{2} \quad\Rightarrow\quad
  B(\lambda) = \frac{\pi h}{2 m_n \gamma l} \frac{1}{\lambda},
\end{equation}
where $h$ is Planck's constant, $l$ is the RF coil thickness, $v$ the neutron
velocity, $m_n$ its mass and $\lambda$ its wavelength.  In a time-of-flight
instrument, the wavelength of a neutron with time of flight $t$ at one component
is given by $\lambda = ht/m_nx$, where $x$ is the component's distance from the
source.  As the MIEZE module was originally developed for operation at a reactor
source, the RF current was not modulated with the time of flight in the present
setup.

As detector we used a circular multi-channel-plate prototype \cite{Tremsin:08},
with a diameter of 40\,mm and only about 15--20\,\% detection efficiency for cold
neutrons, but it provided sufficient time resolution and a very good spatial
resolution (pixel size about 10\,\textmu m), both of which might become very
useful in future tests at higher time resolution.  The detector was always
positioned according to the MIEZE condition
\begin{equation}
  \label{eq:miezecond}
  L_2 = \frac{L_1}{f_2 / f_1 - 1},
\end{equation}
where $L_1$ is the distance between the coils, $L_2$ the distance between second
coil and detector and $f_i$ are the RF frequencies.

The chopper at CG-1D has an opening of 2$^\circ$ and was running at 40\,Hz,
which results in a pulse length of 0.14\,ms.  At the first MIEZE coil,
positioned 0.6\,m from the chopper, this amounts to a wavelength spread of
$\Delta\lambda/\lambda \approx 15\,\%$ at $\lambda=3.5$\,\AA{} and 8\,\% at
$\lambda = 6.5$\,\AA.  At the detector position at 5.07\,m distance, the
$\Delta\lambda/\lambda$ reduced to 1.7\,\% and 0.9\,\%, respectively.

\section{Results}

We report the results from two separate beam sessions, where we implemented
slightly different instrumental parameters.  The goal in the first beam session
was to set up the MIEZE devices for the first time at CG-1D, while the goal in
the second beam session was to perform a first sample measurement.

\subsection{First session}

\begin{figure}[h]
  \begin{center}
    \includegraphics[width=0.85\linewidth]{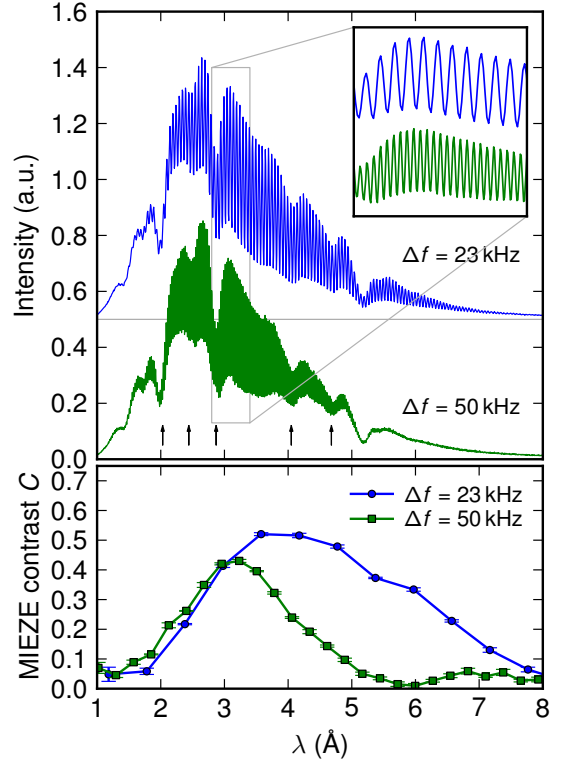}
  \end{center}
  \caption{\label{fig:twofreq}
    Intensity modulation and resulting contrast for the direct beam wavelength
    spectrum for two different MIEZE frequencies, $\Delta f = 23$\,kHz and
    $\Delta f = 50$\,kHz.  The contrast is calculated by binning the spectrum
    into 8 bins per oscillation and fitting a sinusoidal signal to the summation
    of 12 individual oscillations for each point.  The black arrows indicate the
    Bragg absorption edges of Al.
    Note that for better clarity, the spectrum for $\Delta f = 23$\,kHz has been
    shifted by 0.5 on the intensity scale.}
\end{figure}

Figure \ref{fig:twofreq} shows the result of the first test, a MIEZE modulation
on the time-of-flight spectrum of the chopped beam at CG-1D with two different
frequencies.  The absorption edges of aluminum (marked with black arrows) and
silicon (used in the solid-state polarizers) are clearly visible as dips in the
Maxwellian shape of the spectra and were used to calibrate the time-of-flight to
wavelength conversion.  The RF frequencies were 46/69\,kHz and 99/149\,kHz,
respectively, with distances $L_1 = 0.9$\,m and $L_2 = 1.8$\,m to fulfill the
MIEZE condition.  For calculating the MIEZE contrast in these and all other
spectra, the spectrum is binned into 8 bins per MIEZE oscillation.  Each
oscillation spans a time-of-flight interval $t = 1/\Delta f$.  To improve
statistics, 8 to 15 individual oscillations are summed before a sinusoidal
signal
\begin{equation}
  I(n) = B + A\sin(\frac{2\pi}{8}\cdot n + \varphi) \quad\mathrm{with~~}n=1,\dots,8
\end{equation}
with a fixed phase $\varphi$ is fitted to the spectra.  The contrast $A/B$ of
the fitted signal is the MIEZE contrast.  The maximum contrast in our tests was
about 50\,\%, which is most likely due to one of the MIEZE-coil sets having been
damaged during the oversea transport causing static stray fields near the
RF-coils and therefore interfering with the correct operation of the flippers.

Note that for $\Delta f = 50$\,kHz, the effect of not ramping the RF current
becomes clearly visible: the $\pi/2$ flip is achieved for neutrons of 3\,\AA,
which means that at the same current neutrons of 6\,\AA{} will be flipped by
$\pi$, which makes the contrast vanish completely in our MIEZE mode.  If there
was enough intensity at 9\,\AA{}, one should see a considerable contrast again,
in this case making $3\pi/2$ flips in the coils.

\subsection{Second session}

All the following measurements were performed in a second beam time, using only one set
of RF frequencies of 99/124\,kHz.  The frequency ratio was chosen differently
because with eq. (\ref{eq:miezecond}) it results in a longer distance between
second coil and detector, and allowed us to insert a cryomagnet at the sample
position.

\begin{figure}[h]
  \begin{center}
    \includegraphics[width=0.85\linewidth]{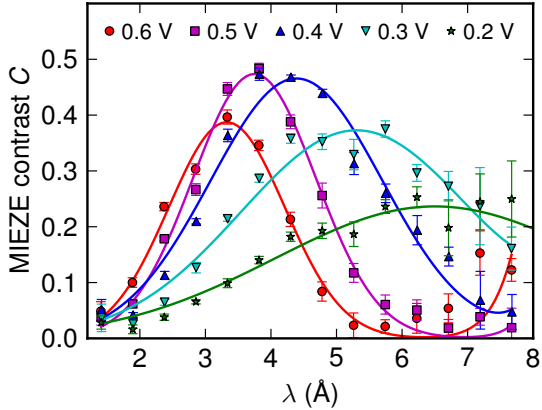}
  \end{center}
  \caption{\label{fig:amplscan}
    Contrast as a function of wavelength in the direct beam for various output voltages applied
    at the RF generator.  The solid lines are guides to the eye.
    The MIEZE frequency was $\Delta f = 25$\,kHz.  Note
    that with eq. (\ref{eq:rfmod}) half the RF field should flip twice the
    wavelength; this does not hold for the output voltage due to
    nonlinearities in the RF amplifier and circuit.}
\end{figure}

To elucidate the effect of RF current ramping, figure \ref{fig:amplscan} shows
the contrast as a function of wavelength in the direct beam for various output voltages in the RF
generators.  Note that on the one hand, towards very short wavelengths the
initial beam polarization decreases rapidly, and therefore the maximum MIEZE
contrast decreases.  On the other hand, for large wavelengths the effect of
stray fields becomes more pronounced, which again influences the maximum
contrast.

\begin{figure}[t]
  \begin{center}
    \includegraphics[width=0.85\linewidth]{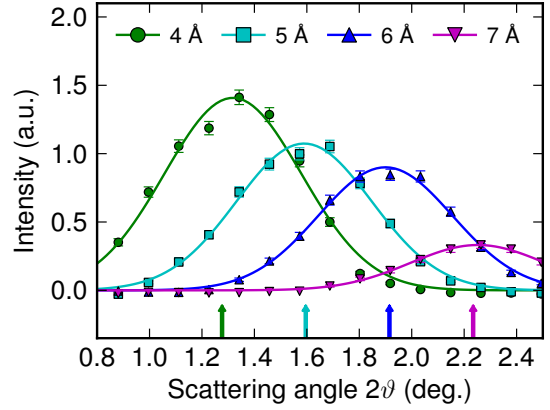}
  \end{center}
  \caption{\label{fig:linescan}
    Intensity of the magnetic MnSi peaks as a function of position on the detector for various
    wavelengths. The detector was moved by 3\,cm out of the direct beam for fulfilling the Bragg condition
    and the intensity was summed over 1\AA{} intervals around the given
    wavelength.  The vertical arrows represent a calculation of the expected
    Bragg positions of the MnSi peak for the mean wavelength of the interval.
  }
\end{figure}

Without changing the MIEZE setup we moved the detector 3\,cm out of the direct
beam in order to measure the neutrons scattered by the helical magnetic order in
a MnSi sample placed inside the cryostat at zero field.  MnSi is a weak
itinerant ferromagnet that below $T_c = 28.9\,$K forms a magnetic spiral with a
pitch $d \sim 180$\,\AA, and magnetic satellite peaks become observable at
$q= 2 \pi/d = 0.035\,$\AA$^{-1}$ in the $\left\langle1\,1\,1\right\rangle$
directions \cite{Lebech:95}.  We aligned the sample such that one of the peaks
falls onto the detector in the small-angle scattering condition: see Figure
\ref{fig:linescan}.  The vertical arrows on the plot are the calculated
positions for the Bragg peak from the helical order using
\begin{equation}
  q \approx \frac{2\pi}{\lambda} \sin(2\vartheta) \quad\Rightarrow\quad
  2\vartheta \approx \sin^{-1}\left( \frac{qht}{2\pi m_nL_0} \right),
\end{equation}
with the position of the sample peak $q = 0.035$\,\AA$^{-1}$, the Planck
constant $h$, the neutron mass $m_n$, the time of flight $t$ between the chopper
and the detector for a distance $L_0=5.066$\,m.  The peak intensity moves
further away from the direct beam for longer neutron wavelengths in order to
maintain Bragg's law.

\begin{figure}[t]
  \begin{center}
    \includegraphics[width=0.85\linewidth]{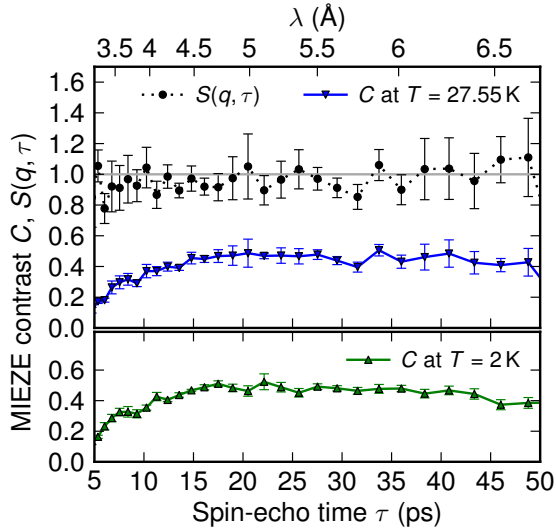}
  \end{center}
  \caption{\label{fig:mnsi}
    The intermediate scattering function $S(q,\tau)$ of the MIEZE signal from
    MnSi at $q=0.035$\,\AA$^{-1}$ is shown for $T=27.55$\,K as measured (blue)
    and after normalization (black) with the instrumental resolution function
    as determined at $T=2$\,K (green).
    The spin-echo times are calculated according to eq. (\ref{eq:miezetime}).}
\end{figure}

For the data analysis, the part of the direct beam still visible on the detector
has been subtracted in the form of a measurement at high temperatures, where no
magnetic signal is present.  The blue curve in Figure \ref{fig:mnsi} shows the
contrast measured in the neutron beam scattered by the MnSi sample at
$T=27.55$\,K at the Bragg peak at $q=0.035$\,\AA$^{-1}$ for different
spin-echo times, corresponding to different wavelengths in the incoming beam
according to
\begin{equation}
  \label{eq:miezetime}
  \tau_{\mathrm{MIEZE}} = \frac{m_n^2}{h^2}\, \Delta f\, L_S\, \lambda^3,
\end{equation}
with the sample--detector distance $L_S=1.035$\,m.

To obtain the intermediate scattering function $S(q,\tau)$, shown as the black
curve in Figure \ref{fig:mnsi}, this data was normalized to the contrast
measured analogously at $T=2$\,K (the green curve), serving as the reference
measurement of the instrumental resolution \cite{Georgii:11}.  It is expected
that this $S(q,\tau)$ is equal to one, since the sample dynamics below $T_c$ are
too slow to be observed at the present time resolution at the Bragg peak
\cite{Georgii:11,Pappas:09}.

To finally assess the field compatibility of the technique beyond what had
previously been demonstrated, we moved the detector back into the direct beam
and looked at the contrast of the MIEZE signal as a function of a horizontal
magnetic field at the sample position.

\begin{figure}[t]
  \begin{center}
    \includegraphics[width=0.85\linewidth]{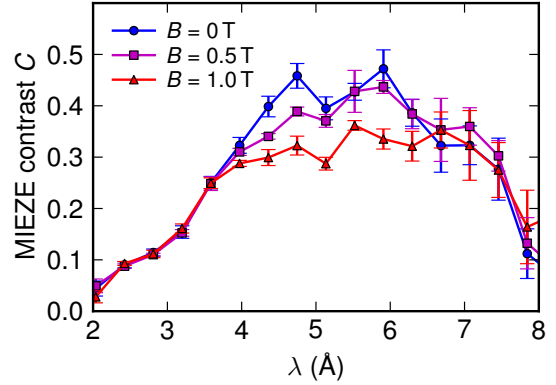}
  \end{center}
  \caption{\label{fig:bfield}
    Contrast of the MIEZE signal in the direct beam for various magnetic fields
    at the sample position.  The small loss of $C$ with increasing field is due
    to the missing coupling coils and the insufficient \textmu-metal shielding.
  }
\end{figure}

Figure \ref{fig:bfield} confirms that the MIEZE technique works with unshielded
horizontal magnetic fields of up to 1\,T.  The observed drop in contrast to
about 70\,\% of its zero-field value we attribute to the fact that we performed
this experiment without coupling coils and with a low-quality \textmu-metal
shielding around the MIEZE coils.  Nevertheless the results show that MIEZE
works even under unfavourable stray field conditions.

\section{Concluding Remarks}
The effort to establish MISANS as a measurement technique for TOF applications
is on the way and this paper documents important results along the path: the
experiment establishes that MISANS can measure samples on pulsed neutron
beamlines and the measurement of a MIEZE-signal in the direct beam with a 1\,T
horizontal field at the sample region demonstrates that this is a most promising
technique for measurements of slow dynamics in high magnetic fields.

In the future, the setup could be extended to develop a MIEZE insert available
as a standard option at several instruments at the SNS or at a dedicated
MIEZE-SANS instrument at HFIR.  It is also planned to propose a MISANS-type
instrument for the European Spallation Source (ESS).  Here, the instrument can
benefit from the extended wavelength range, since the relaxed requirements on
wavelength resolution typical for SANS are also acceptable for MIEZE-SANS.  This
means that similar gain factors can be expected.

\section{Acknowledgements}
We acknowledge the support of Ian S. Anderson for making these tests possible at
HFIR.  We are grateful for the help of Mike Fleenor and the staff at HFIR during
the beam times.  We also acknowledge very helpful discussions with Wolfgang
Häußler and the technical support of Reinhard Schwikowski at the FRM~II.

This work was funded by ONRL, the U.S. Department of Energy, BES-Materials
Science, under Contract DE-AC02-06\-CH117, 
and by the German BMBF under ``Mitwirkung der Zentren der Helmholtz Gemeinschaft
und der Technischen Uni\-ver\-sit\"at M\"un\-chen an der Design-Update Phase der
ESS, F\"or\-der\-kenn\-zeichen 05E10WO1.''


\end{document}